\documentclass[aps,prl,twocolumn,showpacs]{revtex4}
\usepackage{amssymb}

\usepackage{graphicx}

\begin{document}


\title{Multiple Magnetization Peaks and New Type of Vortex Phase Transitions in Ba$_{0.6}$K$_{0.4}$Fe$_2$As$_2$}

\author{Bing Shen$^1$, Huan Yang$^2$, Bin Zeng$^1$,  Cong Ren$^1$, Xibin Xu$^2$ and Hai-Hu
Wen$^{1,2}$$^{\star}$}

\affiliation{$^1$National Laboratory for Superconductivity,
Institute of Physics and Beijing National Laboratory for Condensed
Matter Physics, Chinese Academy of Sciences, P.O. Box 603, Beijing
100190, China}

\affiliation{$^2$ Center for Superconducting Physics and Materials,
National Laboratory of Solid State Microstructures and Department of
Physics, Nanjing University, Nanjing 210093, China}

\begin{abstract}
Magnetization and its relaxation have been measured on
Ba$_{0.6}$K$_{0.4}$Fe$_2$As$_2$ single crystals with T$_c$ = 39 K.
The magnetization hysteresis loops (MHLs) exhibit flux jumps in
the low temperature region, and a second peak-effect in the
intermediate temperature region, especially when the field
sweeping rate is low. Interestingly a third magnetization peak can
be easily observed on the MHLs in the high temperature region.
Further analysis find that the first magnetization peak is very
sharp, which is associated with the strong vortex pinning. However
the first dip of the MHL corresponds to a moderate relaxation
rate, then a second peak appears accompanied by a vanishing vortex
motion. Finally a third magnetization peak emerges and the vortex
motion becomes drastic beyond this threshold. The novel features
accompanying the second magnetization peak suggest a new type of
vortex phase transition.

\end{abstract}

 \maketitle
Investigating the vortex phase diagram and its new feature in
unconventional superconductors is very important for both of the
fundamental sciences and potential applications. In cuprate
superconductors, the vortex physics was boosted greatly with the
discovery of many interesting vortex phases and the transitions
between
them\cite{GBaltter,Vinokur,VG,Fisher,Cubitt,Zeldov,Schilling,Beidenkopf}.
Magnetic hysteresis was used as an important tool to probe the
vortex pinning mechanism, vortex dynamics and vortex phase
transitions.\cite{Klein, GBaltter, Beidenkopf}. The second-peak of
magnetization (namely, the increase of magnetization with magnetic
field) was observed in Bi$_2 $Sr$_2$CaCu2O$_{8-\delta}$ (Bi-2212)
and was understood as a disorder induced first order transition
(transition from a quasi-ordered Bragg glass to disordered vortex
glass)\cite{BraggGlass,Zeldov2}. While in a less anisotropic
material, like YBa$_2$Cu$_3$O$_{7-\delta}$ (YBCO), the broad
second-peak of magnetization was called the fishtail effect, which
was explained as due to the crossover from an elastic to plastic
pinning regimes\cite{Abulafia}. Therefore the second-peak effect,
in either the form in YBCO or in Bi-2212, occurs quite frequently
in cuprate superconductors.

In 2008 the discovery of superconductivity above 50 K in iron
pnictides has added a new member in the family of unconventional
high temperature superconductors\cite{Kamihara2008}. The vortex
matter and phase diagram in this new kind materials were
investigated immediately by a variety of
experiments\cite{yanghuan,Prozorov,bingshen,vanderBeek,DouSX,Pramanik,Kalisky}.
For example, the magnetic hysteresis loops have been investigated
and the second-peak (or fishtail) effect was also observed in the
122, 1111 and 11-families showing a very similar feature as that
in YBCO. It was generally understood as a crossover from an
elastic to plastic vortex motion
regimes\cite{yanghuan,bingshen,vanderBeek,Pramanik,Kalisky}.
Additionally, unconventional Meissner effect was observed in
Ba$_{1-x}$K$_x$Fe$_2$As$_2$ and Ba(Fe$_{1-x}$Co$_x$)$_2$As$_2$
indicating a novel field enhanced pairing strength in the iron
pnictide superconductors\cite{prozorov1}. In this Letter, we
report the magnetization and relaxation in optimally doped
Ba$_{0.6}$K$_{0.4}$Fe$_2$As$_2$. Beside the second magnetization
peak, for the first time we observed a third magnetization peak in
high temperature region. The third peak becomes more pronounced
when the relaxation rate is low. The occurrence of three
magnetization peaks is attributed to the co-existence of large
scale strong pining centers and weak collective pinning centers.
The second peak characterized by some novel features may be
categorized as a new type of vortex phase transition.

The high quality single crystal was prepared by the self-flux
method\cite{LuoHQ}. The superconducting (SC) transition
temperature is up to 39 K as determined from the magnetization
measurements. The crystallization and chemical composition of our
sample were checked by X-ray diffraction and energy dispersive
X-ray microanalysis, both show perfect quality. The magnetization
were measured by the sensitive vibrating sample magnetometer
(PPMS-based, Quantum Design)at the vibrating frequency of 40 Hz
with the resolution better than 1$\times$10$^{-6}$emu. The
magnetic field sweeping rate can be varied from 0.5 Oe/s to 700
Oe/s. The magnetic field H is parallel to c-axis of single crystal
during the measurements.

\begin{figure*}
\includegraphics[width=17cm]{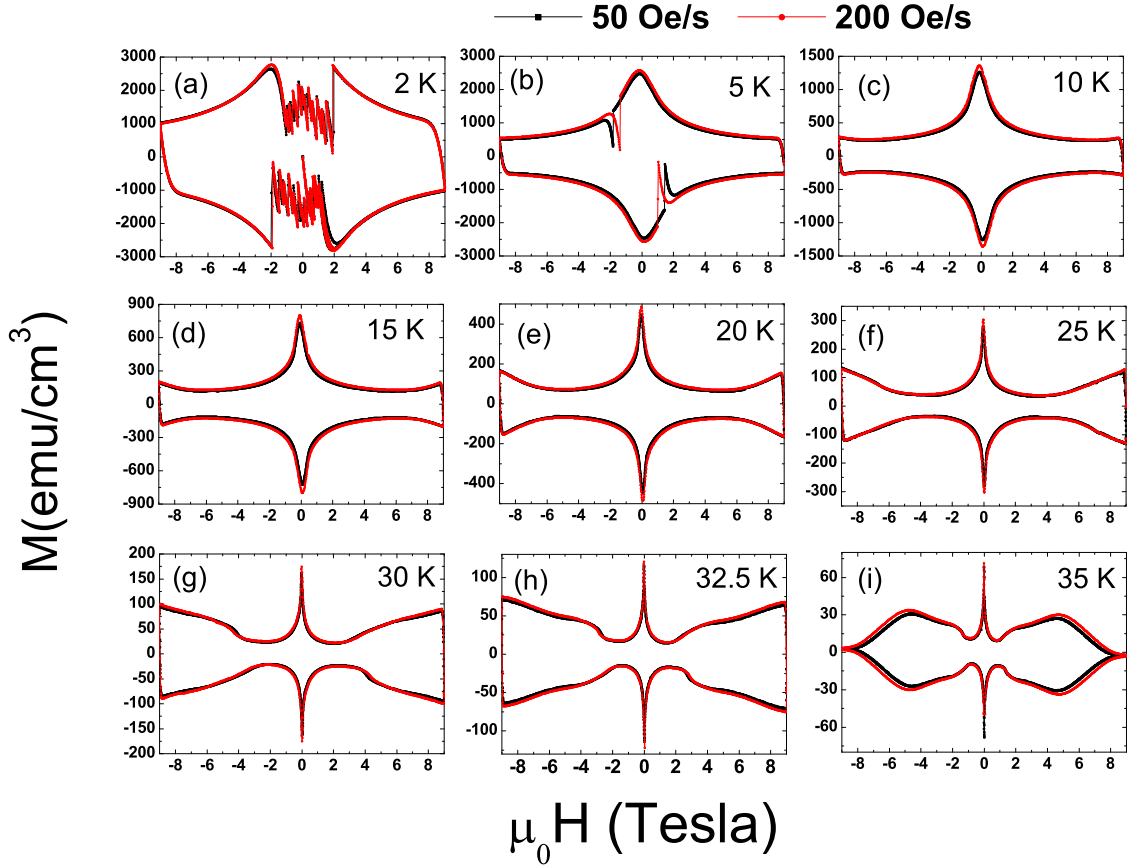}
\caption {(color online) The MHLs measured at different
temperatures with the magnetic field sweeping rates of 50 Oe/s and
200 Oe/s.}
 \label{fig1}
\end{figure*}

Fig.1 shows the magnetization-hysteresis-loops (MHLs) in different
temperature regions with the magnetic field sweeping rates of 50
Oe/s and 200 Oe/s. The symmetric MHL curves suggest that the bulk
pinning instead of surface barrier dominates in the sample. At 2
K, the MHLs exhibit flux jumps (avalanche effect) near zero field
which has been observed in many other
superconductors\cite{Pramanik,zhaozhiwen} and was attributed to
the magneto-thermal instabilities\cite{FluxjumpProzorov}. Above 5
K, a magnetization peak appears near the zero field in the MHLs
and it evolves into a sharp structure at high temperatures. This
sharp central magnetization peak may be understood as an evidence
of strong pinning\cite{vanderBeek, bingshen}. When the magnetic
field is swept back to zero, the Bean critical state model would
assume a flat MHL near zero field if J$_c$(H) is a constant with
J$_c$ the critical current density. The sharp magnetization peak
may suggest that the critical current density is getting stronger
when the field is approaching zero. In moderate temperature
region, after the first central peak, the magnetization increases
with the magnetic field leading to the so-called second-peak (SP)
effect. The SP shifts to lower fields with increasing temperature
and becomes more obvious with more slow field sweeping rate.
Finally, a third peak emerges in high temperature region (see the
data at 32.5 K) and it exhibits the similar temperature dependence
of the SP. For iron based superconductors, there has no report
about three characteristic peaks in MHLs.

\begin{figure}
\includegraphics[width=7cm]{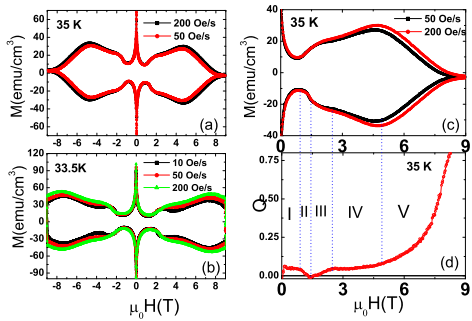}
\caption {(color online) (a),(b):The MHLs with different sweeping
rates at 35 K and 33.5 K respectively. (c),(d): The magnetic field
dependence of the MHL width and the relaxation rate Q at 35 K.}
\label{fig2}
\end{figure}

The MHLs at 33.5 K and 35 K are presented in Fig.2 (a) and (b).
Three characteristic peaks can be observed easily in both curves.
With the fast sweeping rate, the SP exhibits a step-like behavior
(kinky shape), while it becomes more visible (peak shape) with
slower sweeping rate. The position of the SP shifts slightly with
different sweeping rates at the same temperature which causes the
crossing of the MHL curves near SP (shown in Fig.2 (b)). To
further investigate the vortex behavior, we calculated the
corresponding dynamic magnetization-relaxation rate via

\begin{equation}
Q = \frac{\textrm{d} \ln j_s}{\textrm{d} \ln (dB/dt)}=\frac
{\textrm{d} \ln (\Delta M)}{\textrm{d}
\ln(\textrm{d}B/\textrm{d}t)} ,
\end{equation}

where $j_s$ is the transient superconducting current density,
$\textrm{d}B/\textrm{d}t$ is the field sweeping
rate\cite{WenHHPhysicaC1995}. Fig.2(d) shows the magnetic field
dependence of Q at 35 K. Near the valley between the central peak
and SP, a moderate relaxation with Q = 6$\%$ is observed. With
increasing the magnetic field, a clear lowering down of the
magnetic relaxation is observed near the edge of the SP.
Interestingly the relaxation rate Q sometimes shows a negative
value, which we think is not a true effect. It is induced by the
above-mentioned crossing of the MHLs measured at different
sweeping rates, indicating a very slow relaxation rate near the
SP. Such low Q-value and the crossing effect of the MHLs have
never been observed in other superconductors, this may suggest the
very damped vortex motion and strong pinning. We thus categorize
the second peak here as a new type vortex phase transition.

Further increasing the magnetic field, Q rises to a moderate value
again and a plateau of Q can be observed below the third peak (TP)
field in MHL. In high field (beyond the TP) region, the Q rises up
drastically and the magnetization decreases quickly with
increasing the field, suggesting that the vortex dislocations are
proliferated greatly and the shear modula $C_{66}$ drops down
quickly. The vortex dynamics near the third peak is actually quite
similar to the second-peak observed either in YBCO or Bi-2212.

\begin{figure}
\includegraphics[width=8cm]{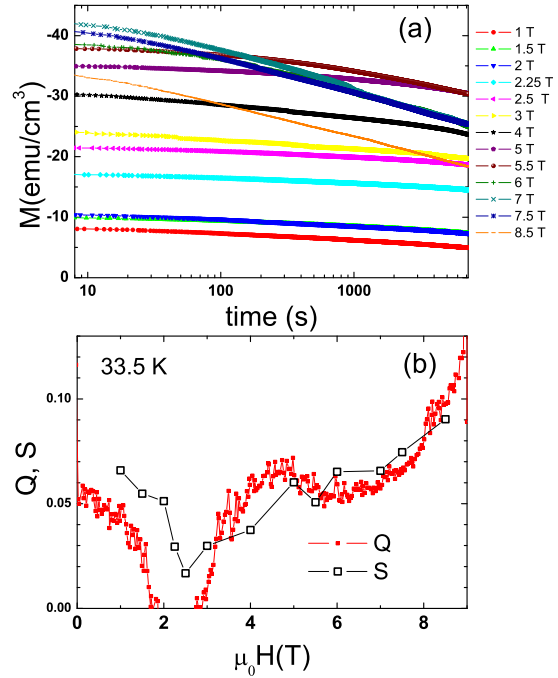}
\caption {(color online) (a) Time dependence of the
non-equilibrium magnetization (M) at 33.5 K. (b)The magnetic field
dependence of the relaxation rate S and Q at 33.5 K.  }
\label{fig3}
\end{figure}

In order to further investigate the vortex behavior and check the
relaxation rate derived from the dynamic relaxation method, we
also did the traditional magnetization-relaxation measurements on
this sample. The time (t) dependence of the non-equilibrium
magnetization (M) at 33.5 K are shown in Fig.3(a). It shows that,
the slow time decay of non-equilibrium magnetization is observed
near 2.5 T which is exactly corresponding to the SP in MHL. Based
on the data of M vs. t, we get the traditional magnetization
relaxation rate defined as S = -d$ln(\mid M \mid)/dlnt$ in
the time window of 100 s to 7200 s. Fig.3(b) shows the magnetic
field dependence of both S and Q at 33.5 K. The data acquired by
the two different methods agree with each other very well. The
abnormal minimum corresponding to the SP can be observed both in
the data of S(H) and Q(H).

Concerning the second peak effect in Bi-2212, different
explanations were given. It was regarded as a consequence of phase
transition between the low field Bragg glass and the high field
vortex glass, which is supported by neutron diffraction, Hall
probe, magneto-optical imaging technique and other
measurements\cite{Khaykovic,vanderBeek1,Gaifullin,YMWang}. It
shows that the SP in MHL has strong time dependence, implying a
dynamical character (transient vortex profile) of the SP effect
\cite{Goffman,Correa,lishiliang,Kalisky1}. In our measurements,
the similar time dependence of SP is found: more slower sweeping
rate, more obvious the SP effect (shown in Fig.2 (b)). This may be
understood in the similar way: large scale strong pinning centers
separate the vortex system into domains, between the domains the
vortex dynamic is dominated by the small scale or point-like
disorders. In Fig.2(d), we separate the vortex dynamics into five
regions due to the presence of the strong and point-like pinning
centers. Region-I near the central peak is characterized by the
strong pinning to very diluted vortices. In region-II, the vortex
number is getting more than that of the large scale pining
centers, therefore the relaxation rate is moderate. When it is in
the SP region, both the large scale and the point-like disorders
behave as effective pinning centers and the vortices may be highly
entangled, therefore the relaxation rate is very low. A further
increase of the magnetic field will make the vortex system more
and more dense. It is in this region (region-IV), the vortex
entity between the large scale pinning centers starts to shear,
leading to a finite relaxation. But the dislocations are still
quite limited and the critical current density are still growing
up. Beyond the TP, large number of dislocations are proliferated
and plastic motion occurs. Therefore the multiple magnetization
peaks in the MHL are induced by cooperative interactions between
the vortices, the large scale pinning centers and the point-like
disorders.

\begin{figure}
\includegraphics[width=8cm]{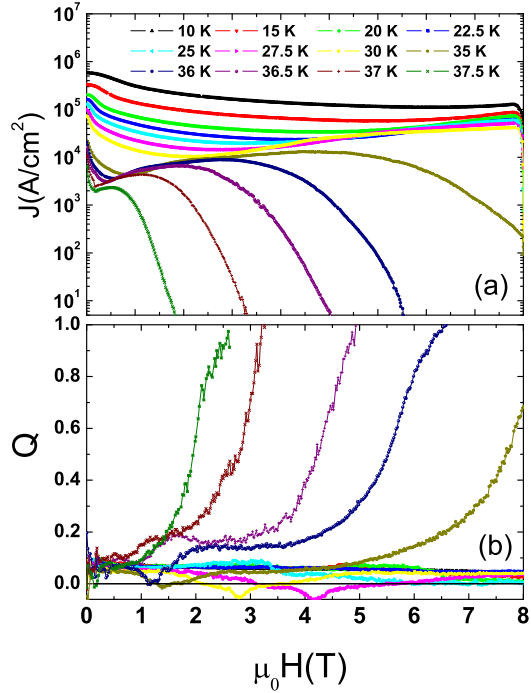}
\caption {(color online) The magnetic field dependence of the
superconducting current density and Q from 10 K to 37.5 K }
\label{fig4}
\end{figure}

Based on the Bean critical state model\cite{Bean}, we calculate
the superconducting current density and shown in Fig.4(a), the
corresponding relaxation rate Q are shown in Fig.4(b). The strong
temperature dependence of the dip in Q(H) corresponding to SP can
be observed. In Bi-2212 the SP shows weak temperature dependence
in low field region and disappear at the melting line. This
validates the conclusion that the SP in Bi-2212 in induced by the
quenched disorder. But in our measurements, the unique temperature
dependence of the SP is observed which suggests a different
origin\cite{Radzyner}. In some experiments on YBCO, the so-called
third magnetization peak was also reported. Later on it was
discovered that this was induced by the vortex channelling along
twin boundaries.  This effect is thus sensitive to the angle
between the orientation of twin boundaries and applied
field.\cite{Oussena} We have looked into this possibility by
measuring the MHLs by fixing the angle between the magnetic field
and c-axis constant, but with different in-plane angles respect to
the field. We found the same results for all different in-plane
angles. The TP accompanied by the instant upturn in Q shifts to
the low field with increasing temperature, which is similar to the
SP observed in YBCO, and some iron based superconductors, such as
the Ba(Fe$_{1-x}$Co$_x$)$_2$As$_2$ and
LiFeAs.\cite{PRADHAN,Abulafia,bingshen,Pramanik}. Below the TP,
the Q keeps a relative low value which suggests the elastic motion
of vortices.

\begin{figure}
\includegraphics[width=9cm]{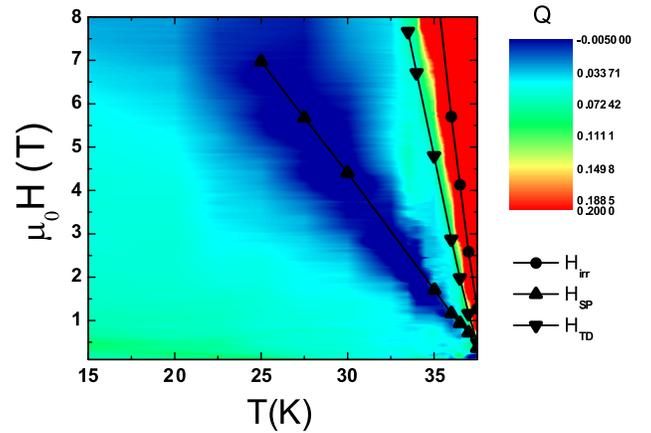}
\caption {(color online) The vortex phase diagram of
Ba$_{0.6}$K$_{0.4}$Fe$_2$As$_2$. The contour with different colors
are correspoding to different relaxation rates Q. The second peak
H$_{SP}$, the third peak H$_{TP}$ and the irreversible line
H$_{irr}$ are derived from the characteristic points of MHLs. }
\label{fig5}
\end{figure}

Fig.5 shows the vortex phase diagram based on the Q values. The
very low relaxation can be observed in a narrow region (the blue
region shown in Fig.5) which is corresponding to the SP. The
third-peak field H$_{TP}$ indicates the crossover from the elastic
motion to plastic motion of vortices. In iron based
superconductors, the disordered vortex structure in a large scale
was observed in local measurements, which suggests strong pinning
effect.\cite{Kalisky,Inosov,Demirdis,Lshan} In our measurements,
the multiple magnetization peaks and the novel vortex dynamics can
be interpreted as cooperative interactions between the vortices,
the large scale pinning centers and weak but dense point-like
disorders. The large scale pinning centers are probably induced by
the domain walls between the superconducting and the fluctuating
antiferromagnetic regions. The point-like disorders are given by
the dopants, like the random distributions of Ba/K. It would be
very interesting to check whether the proposed two different kinds
of the pinning centers, namely the large scale strong pinning
centers, and the point-like disorders are really present and play
the important role for the vortex pinning.

In summary, for the first time, three characteristic magnetization
peaks are observed in Ba$_{0.6}$K$_{0.4}$Fe$_2$As$_2$
superconductors. The second peak-effect is observed in the
intermediate temperature region which is accompanied by
diminishing vortex motion. In high temperature and high field
regions, a third magnetization peak emerges and the vortex motion
is getting drastically. The abnormal MHLs with three distinct
magnetization peaks can be understood by the model with large
scale strong pining and weak collective pinning centers.

\begin{acknowledgments}
We appreciate the discussions with Yosi Yeshurun, Ruslan Prozorov
and Z. Tes$\check{a}$novi$\acute{c}$. This work is supported by
the NSF of China, the Ministry of Science and Technology of China
(973 projects: 2011CBA00102)
\end{acknowledgments}

$^{\star}$ hhwen@nju.edu.cn

\end{document}